# Diameter dependence of the temperature dynamics of hot carriers in photoexcited GaAsP nanowires

Aswathi K. Sivan[1], Lorenzo Di Mario[2], Yunyan Zhang[3], Daniele Catone[2], Patrick O'Keeffe[4], Stefano Turchini[2], Valentina Mussi[1], Huiyun Liu[3], Faustino Martelli[1]

[1]*Istituto per la Microelettronica e i Microsistemi (IMM), CNR, I-00133, Rome, Italy*

[2]*Istituto di Struttura della Materia-CNR (ISM-CNR), Division of Ultrafast Processes in Materials (FLASHit), Area della Ricerca di Roma Tor Vergata, Via del Fosso del Cavaliere 100, 00133 Rome, Italy.*

[3]*Department of Electronic and Electrical Engineering, University College London, London WC1E 7JE, United Kingdom.*

[4]*Istituto di Struttura della Materia-CNR (ISM-CNR), Division of Ultrafast Processes in Materials (FLASHit),  Area della Ricerca di Roma 1, 00015 Monterotondo Scalo  Italy*

Abstract

Semiconductor nanowires (NWs) often present different structural and opto-electronic properties than their thin-film counterparts. The thinner they are the larger these differences are. Here, we present femtosecond transient absorbance measurements on GaAs$_{0.8}$P$_{0.2}$ NWs of two different diameters, 36 and 51 nm. The results show that thinner NWs sustain a higher carrier temperature for longer times than thicker NWs. This observation suggests that in thinner NWs, the build-up of a hot-phonon bottleneck is easier than in thicker NWs because of the increased phonon scattering at the NW sidewalls, which facilitates the build-up of a large phonon density. Moreover, the important observation that the carrier temperature in thin NWs is higher than in thick NWs already at the beginning of the hot carrier regime suggests that the phonon-mediated scattering processes in the non-thermal regime play a major role at least for the carrier densities investigated here ($8 \times 10^{18} \div 4 \times 10^{19}$ cm$^{-3}$). Our results also suggest that the phonon scattering at crystal defects is negligible compared to the phonon scattering at the NW sidewalls.

*E-mail address: faustino.martelli@cnr.it

1. Introduction

Semiconductor nanowires (NWs) are quasi-one-dimensional structures with tunable optical and electronic properties. [1–5] Due to their large surface area to volume ratio, NWs have excellent photonic properties that are suitable for use in novel light-emitting diodes [6,7], lasers [8], solar cells, [9,10] and photocatalysts. [11–13] Apart from being a technologically versatile material system, they are ideal for manipulating and studying several fundamental and novel physical phenomena at lower dimensions. For example, heat flow is a fundamental energetic process in materials that shows a clear diameter dependence in NWs, due to their constraining quasi-one-dimensionality. As first reported for Si and SiGe NWs by Li and coworkers, the thermal conductivity decreases as the diameter of the NW shrinks. [14,15] In particular, in those works it was shown that the thermal conductivity decreases for diameters smaller than 100 nm and changes its temperature dependence. Swinkels *et al.* [16] reported the diameter dependence of thermal conductivity in InAs NWs with diameters in the 40-1500 nm range and observed a strong reduction in the thermal conductivity for diameters below 50 nm. The inverse diameter dependence of thermal conductivity in NWs can be ascribed to the boundary scattering of the phonons. [14,17] As the surface-to-volume ratio increases with the reduction of the diameter, the phonon scattering at the NW surface is indicated as the main reason for the thermal conductivity reduction, with relevant contributions from the surface roughness. [18,19] Their studies suggest that the diameter shrinkage might induce changes in the phonon-carrier interaction, e.g., in the hot-carrier cooling. During photoexcitation in a semiconductor, photons with energies greater than the bandgap excite electrons from the valence band to the conduction band, creating a non-equilibrium distribution of electrons (holes) in the conduction (valence) band. Initially, these carriers undergo rapid thermalization to form a Fermi-Dirac thermal distribution characterized by an effective carrier temperature ($T_c$) that is usually higher than the lattice temperature, $T_L$, leading to the use of the term hot-carriers. The thermalization process is induced by rapid carrier-carrier scattering, often indicated as the main scattering process in this regime, carrier-phonon scattering and, depending on material and excitation energy, intervalley scattering. [20] The characteristic time for it to lead to a thermal distribution of hot-carriers has been indicated to be on the tens of femtoseconds (fs) timescale, as also recently reported for InAs NWs. [21]

The knowledge of the dynamics of hot-carriers is important for the development of many optoelectronic devices. Depending on the device type a more rapid or a slower cooling of the hot-carriers would be preferred. For example, a faster cooling is needed in semiconductor lasers based on interband transitions, [22] while a slower cooling may be beneficial in intraband infrared lasers [23] or photodetectors. [24] In recent years a fierce debate has developed about the possible exploitation of hot-carriers in photovoltaic cells, [25] for which a slow cooling of the hot-carriers is suggested to be beneficial. [26]

After the thermalization, the hot-carriers relax to the band edge by cooling down to $T_L$ in the next few picoseconds. This relaxation takes place through carrier-phonon interactions, with the emission of longitudinal optical (LO) phonons. [27] After this carrier relaxation, electrons and holes recombine radiatively or non-radiatively. The rate of cooling of the hot-carriers depends on the carrier-phonon interaction. We then expect that when the diameter of the nanowire reduces, perturbing the phonon scattering at the NW sidewalls, the carrier-phonon interaction changes leading to changes in the hot-carrier cooling rates.

Though there are some indications in the changes in carrier-phonon scattering rates as a function of diameter, through electrical measurements, [16] there are only a few reports on the optical study of this phenomenon. [28–30] In particular, Tedeschi *et al*. [28] have shown that steady-state photoluminescence (PL) spectra of InP and GaAs NWs, excited by a continuous-wave laser, present a high-energy tail of the PL band that is explained by a $T_c$ higher than $T_L$. It is shown that $T_c$ increases as the NW diameter decreases. The presence of the high-energy tail is a clear indication of the presence of long-lived hot-carriers in InP and GaAs NWs of small diameters. Through time-resolved PL measurements, Yong *et al.* [29] have found that 50 nm wide wurtzite InP NWs sustain hot-carriers longer than InP NWs with a diameter of 160 nm. However, in that work the authors have attributed the changes in cooling rate not to the smaller diameter but to the larger density of stacking fault or density of zincblende inclusions observed in the thinner wurtzite NWs.

As written above, Tedeschi *et al.* [28] and Yong *et al.* [29] have investigated the presence of long-lived hot-carriers employing steady-state and time-resolved PL, respectively. In the present work, we have used transient absorbance (TA) measurements, which give a direct picture of the hot-carrier dynamics in the energy range above the direct bandgap of GaAsP. In particular, by femtosecond pump-probe spectroscopy, we directly measure the time dependence of $T_c$ in GaAsP NWs of two different diameters, pointing out the importance of the phonon-mediated scattering processes in the non-thermal regime of the carrier relaxation process.

2. Experimental details and samples characterization.

The self-catalyzed ternary alloy NWs of the nominal compound composition GaAs$_{0.8}$P$_{0.2}$ were grown by molecular beam epitaxy (MBE) on Si substrates. For the alloy composition used, the NW samples have a direct bandgap, see below. The details of the growth can be found in Zhang et al. [31] Figures 1(a) and 1(b) show the scanning electron microscopy images of the two samples used here. The first image (a) shows thin NWs with an average diameter of 36 nm, while the second image (b) shows sample with thicker NWs with an average diameter of 51nm.

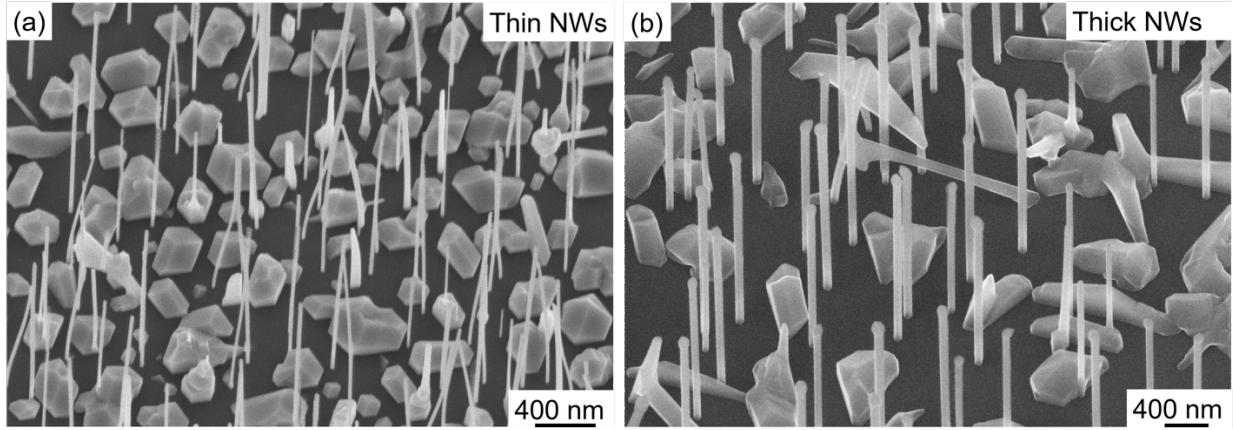

**Figure 1**. 35° tilted scanning electron microscopy images of (a) thin (average diameter of 36 nm) and (b) thick (average diameter of 51 nm) GaAsP NWs.

The dynamics of the $T_c$ in the NWs were probed using a fast-transient absorbance spectroscopy (FTAS) in pump-probe configuration in a femtosecond TA spectrometer (IB Photonics FemtoFrame II). The experiments were carried out with the help of a laser system consisting of Ti: Sapphire oscillator and a regenerative amplifier giving an output of 4 mJ, 35 fs wide laser pulses at 800 nm with a repetition rate of 1 kHz. Part of the output of the amplifier passes through an optical parametric amplifier (OPA) with a tunable output, used as the pump to excite the samples. In the present case we have used the wavelength of 430 nm (2.88 eV). The response function of the system is about 50 fs. A supercontinuum white light was used as a probe in the visible region of the spectrum. The white light supercontinuum is generated in the visible region by focusing 3 µJ of 800 nm into a rotating $CaF_2$ crystal inside the spectrometer. The output of the measurements is the differential absorbance, $\Delta A$, defined as the difference between the absorption of the probe measured after the photoexcitation by the pump and the unperturbed absorption of the probe at the same energy. More detailed explanation of the pump-probe setup can be found elsewhere. [32]

In contrast to the possibility of direct measurements when the NWs are grown on transparent substrates, [33,34] because our GaAsP NWs are grown on Si, which is opaque to the light used in the pump-probe experiments, it was necessary to carry out the optical measurements on NW ensembles transferred onto a transparent quartz substrate by mechanical rubbing. The measurements were performed at 77 K with pump fluences of 260 µJ/cm$^2$ and 52 µJ/cm$^2$. For these fluences, we roughly estimate initial excited carrier densities of $4 \times 10^{19}$ cm$^{-3}$ and $8 \times 10^{18}$ cm$^{-3}$, respectively. [35] As no absorption coefficient is available for GaAs$_{0.8}$P$_{0.2}$, to evaluate the photoexcited carrier density we have taken the value found for GaAs 1.2 eV above the band gap energy in the work by D.E. Aspnes and A.A. Studna. [36] In our experiments, 1.2 eV approximately corresponds to the difference between the pump energy and the GaAs$_{0.8}$P$_{0.2}$ band gap.

Prior to the TA measurements, the samples were characterized by PL and Raman spectroscopy to assess the structural and optical quality of these NWs. The steady-state PL measurements were performed at different temperatures exciting the as-grown samples with a laser diode at 405 nm. The Raman measurements were carried out on as-grown samples with a DXR2xi Thermo Fisher Scientific Raman Imaging Microscope with laser excitation at 532 nm.

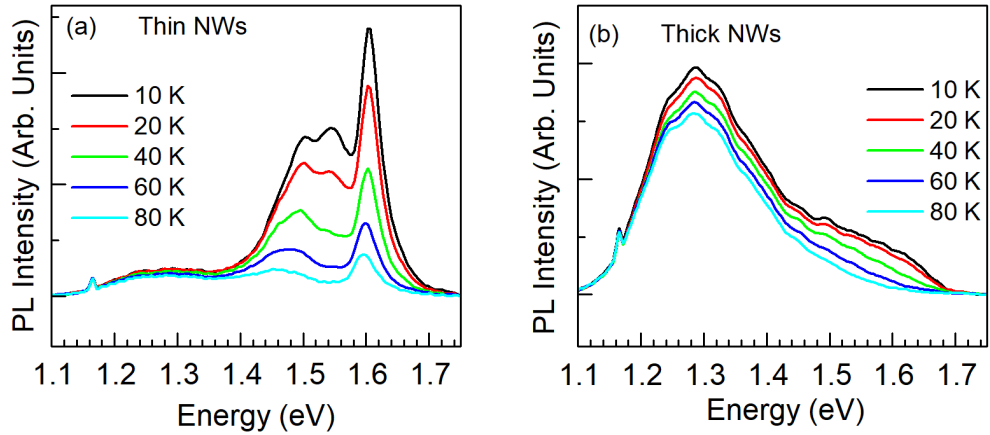

**Figure 2.** Temperature dependent PL of (a) thin NWs and (b) thick NWs.

Figure 2 shows the temperature dependent PL spectra of the two NW samples. The spectrum of thin NWs (Figure 2(a)) is composed of a dominant, narrow near-band edge (NBE) recombination at 1.632 eV and by a number of low-energy emissions related to impurity- and defect-related transitions. In the spectrum of thick NWs (Figure 2(b)), the NBE emission appears as a shoulder of a broader, defect-related low-energy band. The PL spectra suggest that the optical quality is better in thin NWs. The apparent lower optical quality of the thick NWs is confirmed by the Raman spectra shown in Figure 3. The relatively large Stokes shift observed in the PL with respect to the value of the absorption bleaching peak, about 100 meV, can be explained by the alloy composition fluctuations present in the ternary NWs. Indeed, the absorption energy will reflect the average alloy composition (see below), while PL will favor low-energy transition in lower P-content regions of the NWs.

Figure 3(a) and Figure 3 (b) show the Raman spectra of thin and thick NWs, respectively. For both samples, the Raman spectra consist of three peaks due to the GaAs-like TO and LO phonons, at lower energy, while the high-energy signal is attributed to light scattering with the GaP-like LO phonon. The relative energies are 269.1 cm$^{-1}$, 287.8 cm$^{-1}$ and 356.8 cm$^{-1}$ for the thin NWs, and 269.1 cm$^{-1}$, 286 cm$^{-1}$ and 355 cm$^{-1}$ for the thick NWs. Thin NWs show narrower, well defined, and more intense peaks as compared to thick NWs for the same excitation conditions. A small red shift is also observed in thick NWs for some peaks with respect to thin NWs probably due to a higher defect density. [37] These Raman spectra confirm that the thin NW sample is of better crystalline quality than the thick NW sample.

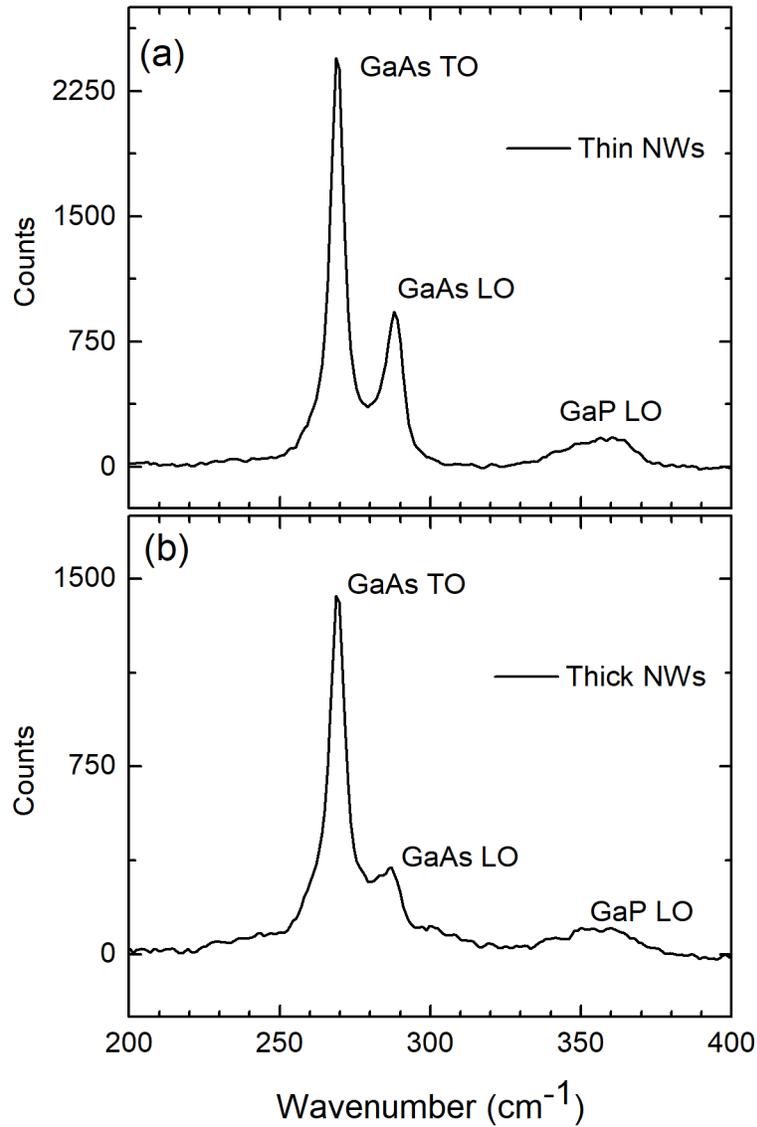

**Figure 3.** Raman spectra of as-grown (a) thin and (b) thick NW samples.

3. Results and discussions.

Figure 4 presents the results of the FTAS measurements at 77 K for the GaAsP NWs. Figures 4(a) and 4(b) show the 2D false-colormap of the ΔA for thin and thick NWs, respectively. The signal intensity, given by the color scale, is reported as a function of probe energy (x-axis) and of the pump-probe time delay (y-axis). The 2D false-colormaps show negative signals at energies above 1.6 eV, indicating a reduction in absorbance, called absorption bleaching. The reduction of absorption of the probe results from the carrier (de)population of the (valence) conduction band due to photoexcitation by the pump. The spectral analysis shows two bleaching peaks. A high intensity peak centered around 1.72 eV and a second, less intense peak centered around 2.03 eV. The peak at 1.72 eV corresponds to the absorption bleaching at the energy of the direct transition at the Γ point, while the peak at 2.03 eV corresponds to the Γ-X (indirect) transition, see below for the discussion about these assignments. Between the two transitions and above the Γ-X transition, a broad bleaching tail extends towards high energies. In detail, Figure 4(c) shows the ΔA of both thick and thin NWs at a time delay of 300 fs. In agreement with what is expected from the previous characterizations, the absorbance bleaching, especially at 2.03 eV, is narrower for thin NWs than for thick NWs.

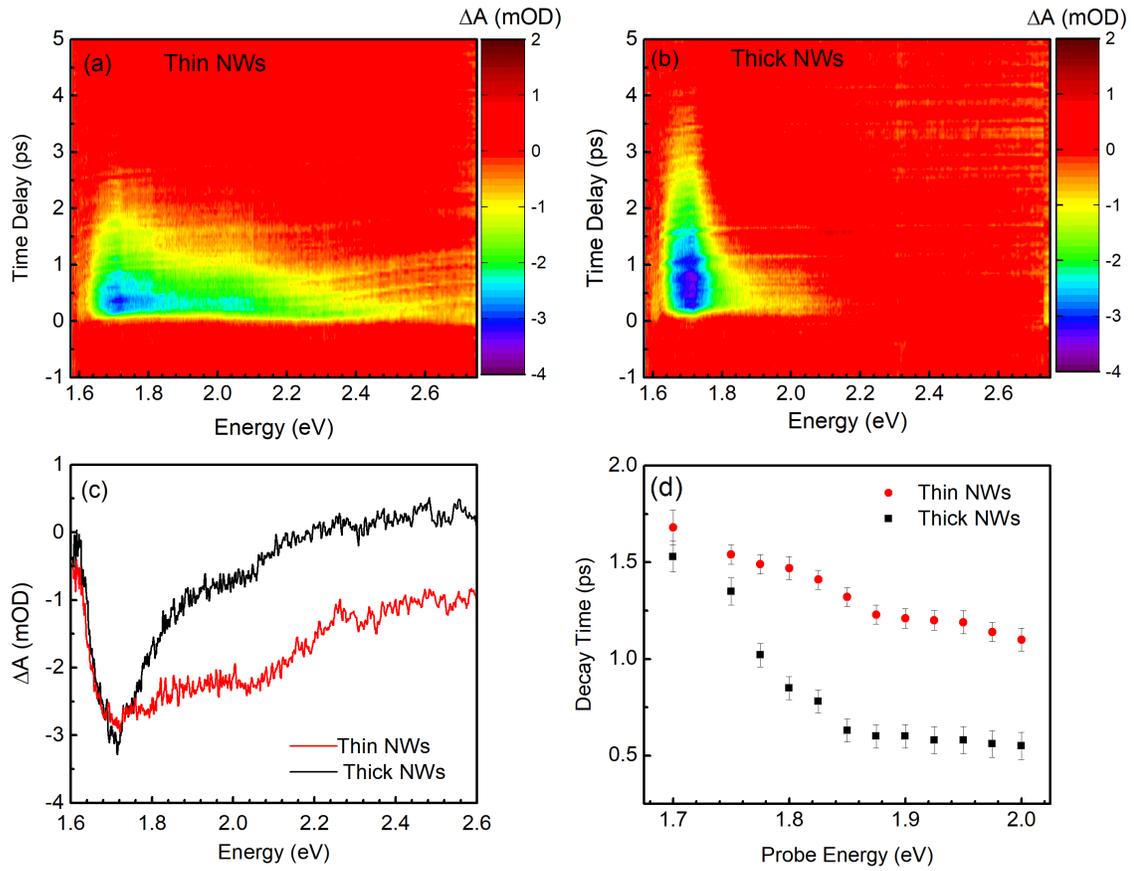

**Figure 4.** 2D false colormap of ΔA for: (a) thin NWs and (b) thick NWs, when they were pumped at 2.88 eV ($E_p$) with a fluence of 260 μJ/cm.² The signal intensity, given by the color scale, is reported as a function of probe energy (x-axis) and of the pump-probe time delay (y-axis); (c) TA spectra for thick NWs (black) and thin NWs (red) at a time delay of 300 fs; (d) Decay time constants of the absorbance bleaching signal estimated at different probe energies for thick (black) and thin (red) NWs.

Before describing the carrier dynamics, we will discuss the spectral characteristics of the absorption bleaching. For GaAs$_{1-x}$P$_x$ alloys, the bandgap increases for increasing P content and it changes from direct to indirect character around $x \approx 0.44$. [38,39] Using the calibration curves given in refs. [38] and [39] and the energy of the direct transition at 77 K measured with our FTAS (1.72 eV), we estimate the alloy composition of our NWs to be x≈0.17, in very good agreement with the nominal value given by the growth parameters (x≈0.2). The indirect bandgap relative to this alloy composition is then expected[38,39] to be at 2.01 eV which is in excellent agreement with the energy value found in our FTAS measurements (2.03 eV) for the high energy peak. This agreement supports our attribution of the high-energy peak to the Γ-X (indirect) transition. As indirect band gap transitions do not show a sharp absorption edge, a clear absorption bleaching peak is generally not observed for indirect transitions, see, e.g., the case of Si. [33] In our case of the ternary GaAsP alloy, the relatively strong absorption bleaching peak observed at the Γ-X (indirect) transition is made possible by the disorder induced by the local potential fluctuations of the alloy composition that break down the translational symmetry of the crystal Hamiltonian. As a result, the disordered crystal potential induces a character mixing between the Γ and the X states that allows no-phonon indirect optical transitions. [40] Above the two peaks a broad energy tail is observed due to the presence of a hot-carrier related signal.

Incidentally, it is worth pointing out that the high sensitivity of FTAS to small changes in the absorbance makes it possible to access electronic transitions otherwise difficult to observe in NWs with steady-state absorption or modulated reflectivity measurements.

The temporal analysis of the TA measurements was carried out by fitting ΔA as a function of the time delay with a single exponential decay function. The results at different probe energies are summarized in Figure 4(d). The absorbance bleaching at 1.72 eV has a decay time constant of ~2.2 ps for both samples. As expected, the decay time decreases with increasing energy because it is related to hot-carriers that relax to low-energy states, and, at all energies, it is shorter for the thicker wires. At 2.03 eV, the absorbance bleaching signal has a decay time of ~500 ± 70 fs for thick NWs and ~ 1.1 ± 0.1 ps for thin NWs. The decay of the bleaching signal is accelerated at higher energies relative to the bandgaps because of the depopulation of excited levels through carrier-phonon scattering that cools the carriers down to the $T_L$. Figure 4(d) gives direct evidence for a faster decay rate of photo-excited hot-carriers in thick NWs compared to thin NWs.

$T_c$ can be determined by fitting the intensity of the high energy tail of $\Delta A$, as shown in Figure 5(a). By extracting $T_c$ at different time delays, it is possible to determine the time taken for the hot-carriers to cool down to $T_L$. The high energy tail of the TA spectra can be fit by a Maxwell-Boltzmann (MB) distribution defined by a carrier temperature $T_c > T_L$, and from this fit $T_c$ at different time delays can be extracted. This was done by performing two separate fits taking the values of the energy of the direct and the indirect transition bleaching signals as the reference energy ($E_0$ in formula 1). As an example of the fitting procedure, Figure 5 (a) shows the MB fitting of the high energy tail with respect to the direct (magenta) and indirect (green) bandgaps of the TA at a time delay of 0.7 ps to extract $T_c$ for thin NWs. The well-known relationship is:

$$I(E) = I(E_0) \exp (E-E_0/k_B T_c) \qquad (1)$$

where I(E) is the modulus of the absorption bleaching value as a function of the probe energy and $k_B$ is the Boltzmann constant.

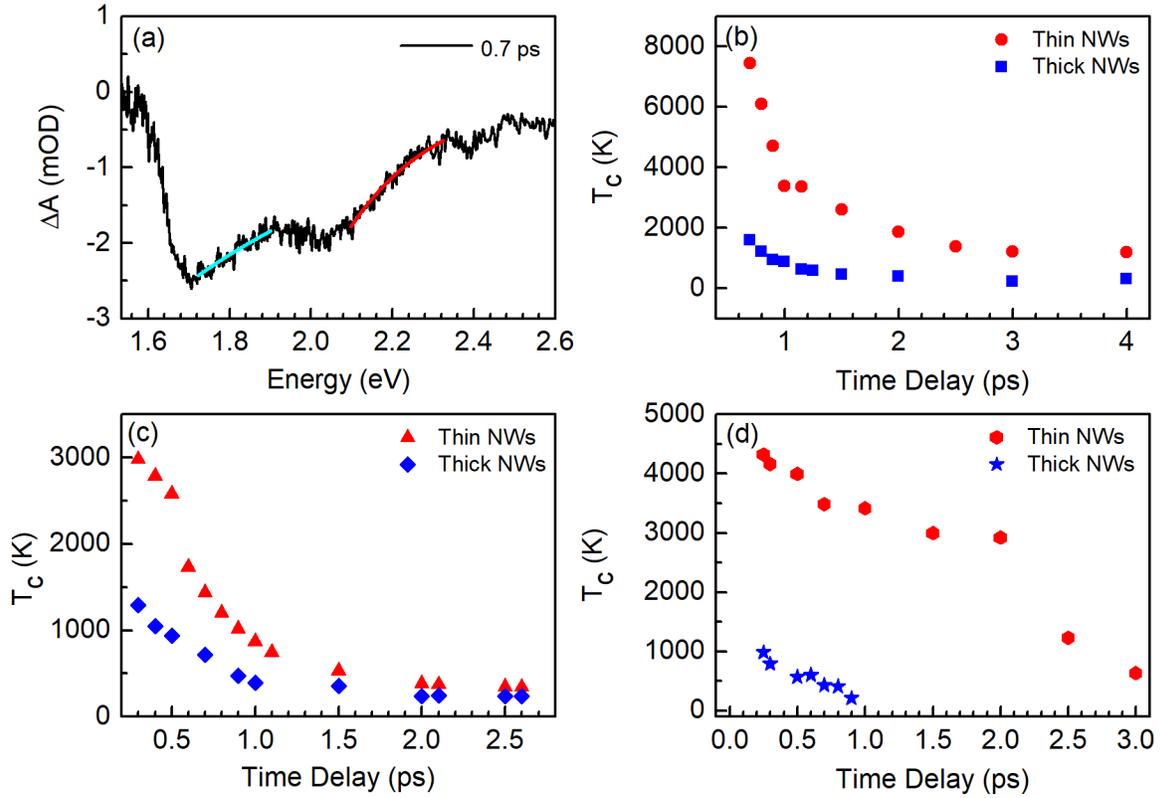

**Figure 5.** (a) ΔA at a time delay of 0.7 ps and a fluence of 260 μJ cm$^{-2}$ for thin NWs along with two Maxwell-Boltzmann function fits to extract the $T_c$ with respect to the direct (magenta) and indirect (green) bandgap (see text for the details); (b) $T_c$ as a function of time delay for thin (red) and thick (blue) NWs with respect to the direct bandgap bleaching peak when they were excited with a fluence of 260 μJ cm$^{-2}$ and (c) 52 μJ cm$^{-2}$; (d) $T_c$ as a function of time delay with respect to the indirect bandgap bleaching for thin (red) and thick (blue) NWs when excited with a fluence of 260 μJ cm$^{-2}$. The values of $T_c$ for thick NWs in (d) are limited due to the poor signal-to-noise ratio obtained at longer time delays.

The reason for performing two separate analyses for the two energy ranges, between the two bleaching peaks and above the indirect transition, is that the density of states and the effective mass of the carriers significantly differ in the two different conduction band minima. [41] It is hence not possible to consider a unique parabolic band approximation and it is known that the band non-parabolicity plays a role in the carrier cooling. [42] The limit of this approach is that the energy of the Γ-X transition is not a thermal equilibrium energy for the electronic system, as electrons on the bottom of the X minimum rapidly decay towards the Γ minimum as indicated by the faster bleaching decay of the signal at 2.03 eV with respect to that at 1.72 eV. In this way the value of $T_c$ deduced at energies above the indirect electronic transition could be underestimated. However, this aspect does not affect the comparison of the carrier dynamics between NWs of different diameter.

Figures 5 (b) and 5(c) show the $T_c$ values in the first 3 ps extracted from the MB fitting of the high energy tail of the TA with respect to the direct bandgap bleaching at 1.72 eV for different time delays for thin NWs (red symbols) and thick NWs (blue symbols), when they were excited using pump fluences of 260 µJ cm$^{-2}$ and 52 µJ cm$^{-2}$, respectively. Instead, Figure 5(d) shows $T_c$ as a function of the time delay for the energy tail above the indirect bandgap, at 2.03 eV, for thin (red symbols) and thick NWs (blue symbols), with the fluence of 260 µJ cm$^{-2}$.

From Figure 5(b), (c), and (d), it is clear that at any given time delay, $T_c$ in thin NWs is always higher than that in thick NWs. This feature is independent of the pump fluence. We have carried out the MB fits for TA signals starting from a time delay of 300 fs to allow for complete carrier thermalization, at the end of which we can define a temperature for the energy distribution of the carriers. The initial $T_c$ obtained for a time delay of 300 fs, shows that the $T_c$ value is higher for thin NWs compared to thick NWs. Therefore, the carrier thermalization leads to an "initial" $T_c$ that is lower in the thick NWs than in the thin ones. The non-thermal regime is characterized by several scattering processes that all contribute to thermalize the carrier population. These include carrier-carrier scattering, carrier-LO/TO phonon scattering and intervalley scattering, with the former often considered as the main scattering mechanism.[20] Because we used a pump energy larger than the band gap minimum at the X point, all three mechanisms are present in our case. The carrier-carrier scattering is dependent on the density of photoexcited carriers, which, for each fluence, is the same in the two types of NWs used in our work. Therefore, this initial difference of Tc in the two types of NWs should be found in the carrier-phonon scattering and in the intervalley scattering that also includes the emission of optical phonons.[20]

The higher $T_c$ in thin NWs persists over the entire delay span for which a reliable fit of the high energy tail of ΔA was possible (4 ps). In the regime of hot carriers, in which they cool down by emitting LO phonons, a quantity that helps understand the cooling rates in the two samples is the analysis of the energy loss rate. The average energy loss rate by LO-phonon emission is given by: [43,44]

$$\frac{dE}{dt} \sim \frac{\hbar\omega_{LO}}{\tau_{in}} \left(\exp\left(-\frac{\hbar\omega_{LO}}{k_B T_c}\right) - \exp\left(-\frac{\hbar\omega_{LO}}{k_B T_L}\right)\right) \qquad (2)$$

The rate of loss of kinetic energy, $J_c$, is also be given by,

$$\frac{dE}{dt} = J_c = \frac{3}{2} k_B \frac{dT_c}{dt} \qquad (3)$$

where $\tau_{in}$ is the characteristic time for the energy loss through the electron-LO-phonon interaction and $\omega_{LO}$ is the LO phonon frequency. The LO phonon energies are 35.46 meV and 45.1 meV for GaAs and GaP modes respectively. [45] When the phosphorous content is low, the GaP phonon modes are less intense, following a linear relation with *x*. [46]

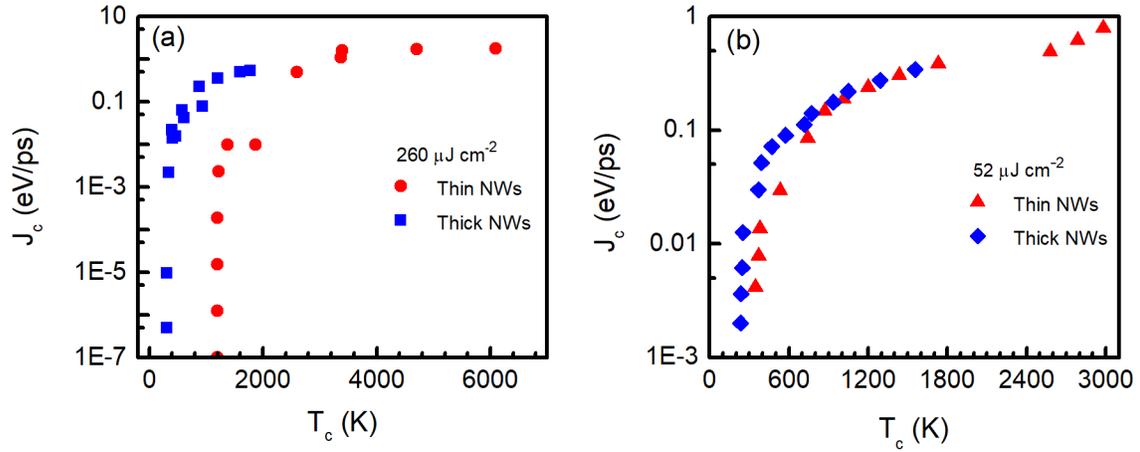

**Figure 6.** $J_c$ as a function of $T_c$ for thin (red) and thick (blue) NWs, with respect to the direct bandgap transition, when excited with a pump fluence of (a) 260 μJ cm$^{-2}$ and (b) 52 μJ cm$^{-2}$, respectively. The samples were excited with a pump at 2.88 eV.

Figure 6 shows the $J_c$ values estimated using formula 3 through the numerical differentiation of the $T_c$ values obtained by the MB fit. The calculations were done with respect to the direct bandgap transition. Figures 6 (a) and (b) show the $J_c$ values when the NWs were excited with a fluence of 260 μJ cm$^{-2}$ and 52 μJ cm$^{-2}$, respectively. The red symbols represent the thin NWs and the blue symbols represent the thick NWs. In Figure 6(a), the $J_c$ values remain high and almost constant above $T_c \sim 2500$ K for thin NWs (red symbols). Below this value of $T_c$, the energy loss rate sharply decreases by several orders of magnitude. Similarly, for thick NWs (blue symbols), the $J_c$ values remain almost constant and high above $T_c \sim 700$ K and drops by several orders of magnitude for lower $T_c$. In Figure 6(b), the $J_c$ values obtained for the low excitation fluence of 52 μJ cm$^{-2}$ are presented. In this case, the $J_c$ values decreases sharply below $T_c \sim 740$ K and $T_c \sim 470$ K for thin NWs (red symbols) and thick NWs (blue symbols), respectively. The initial values of $J_c$ are similar for both the excitation fluences and diameters. In agreement with what is shown above relative to the delay-time dependence of $T_c$, the value of $T_c$, for which we observe the sharp decrease of the $J_c$ value, is higher for the thinner NWs.

The carrier cooling is usually described by an initial fast cooling of the carriers due to the efficient coupling between carriers and the LO-phonons. The LO-phonons must then decay into acoustic phonons [47] to further equilibrate with their environment. Due to rapid cooling of the hot-carriers, a non-equilibrium population of LO-phonons can be then created, building up a hot-phonon bottleneck. [48] The creation of the out of equilibrium population of hot phonons can take place as fast as within 50 fs. [49] The hot phonon bottleneck can be also sustained by the subsequent build-up of non-equilibrium acoustic phonons, that can undergo an acoustic to optical phonon up-conversion process which prolongs the LO phonon lifetime and further slows down the carrier cooling process. [50,51] The acoustic to optical phonon up-conversion is favored in thin NWs where the small diameter limits the acoustic phonon propagation, favoring the phonon-phonon scattering, and limiting the thermal conductivity.

In case of high excitation intensity, the excited carrier density is higher, hence the initial fast cooling creates higher amount of LO phonons as compared to the case of low excitation intensity. This is reflected in the formation of a phonon-bottleneck for higher $T_c$ observed in Figure 6(a) compared to Figure 6(b). A reduction of the hot-carrier cooling rate for reduced dimensionality has been observed in quantum wells [52] and quantum wires [53,54] above some critical carrier density that is always estimated to be below the lowest generated in our GaAsP NWs ($8 \times 10^{18}$ cm$^{-3}$). As explained in the introduction, the thermal conductivity in NWs decreases as the NW diameter decreases, [14] a feature that was explained as due to increased phonon scattering at the NW sidewalls, which is the dominant phonon scattering mechanism. The rate of boundary scattering of phonons increases as the NW diameter decreases. [19] The easier build-up of a hot-phonon bottleneck in quantum wires has been attributed to phonon confinement[51] due to boundary scattering, in analogy to what has been suggested for the diameter dependence of the thermal conductivity in NWs.[14,15,19,20] Moreover, changes in phonon dispersion due, e.g., to zone folding, could modify the selection rules and scattering rate of phonon–phonon scattering.[14]

The increase of the phonon population, because of the hot-carrier cooling, is facilitated in GaAs, and hence in GaAs-based alloys like our $GaAs_{0.83}P_{0.17}$, because the long optical phonon lifetime allows the phonon population to build-up efficiently. [49,55] Therefore, we suggest that an easier build-up of the phonon bottleneck for thin diameters has similar physical origin of the reduced thermal conductivity due to the increased surface scattering and this is likely the reason for the observed reduced carrier cooling in the thinner NWs. As also the $T_c$ measured at very short times is higher for thin wires than for thicker NWs, we suggest that the phonon-surface scattering has also an effect on the efficiency of the carrier-phonon scattering and intervalley scattering in the non-thermal regime and that, at least for the excitation conditions used here, these phonon-related scattering mechanisms play a major role also in the non-thermal regime. We have observed a similar behavior in hybrid lead halide perovskites where we have found that at the end of the thermalization process $T_c$ is higher for smaller grain sizes. [56]

Finally, as Raman and PL measurements suggest that the thinner NWs are of better crystalline quality, our results clearly indicate that phonon scattering at crystal defects, like WZ-ZB stacking faults, is negligible with respect to the scattering at the NW sidewalls surface that increases as the NW diameter decreases and gives rise to the observed slower cooling of the hot-carriers in the thin NWs. This aspect clearly points toward the critical role played by the NW diameter in determining the phonon scattering and hence the carrier-phonon interaction.

4. Conclusion.

We have demonstrated through fast transient absorption measurements of GaAsP NWs of two different average diameters that thin (~36 nm) NWs can sustain hot-carriers for longer times at an elevated temperature compared to thick (~51 nm) NWs under the same excitation conditions. The sustenance of hot-carriers at elevated temperatures for longer times is attributed to an easier build-up of a phonon bottleneck in the thin nanowires as the result of the increased phonon scattering at the NW sidewalls, with the ensuing increase of the LO phonon population because of the photoexcitation of the NWs. The presence of the phonon bottleneck is reflected in a slowing down of the hot-carrier cooling process and in a less efficient carrier-phonon scattering. The observation of a higher carrier temperature in thinner NWs already at the beginning of the hot-carrier cooling suggests that the phonon scattering at the NW sidewall surfaces plays a main role also during the non-thermal regime of the carrier relaxation processes. Finally, our results show that in typical NWs, phonon boundary scattering at defects is negligible with respect to scattering at the sidewall surfaces.

Acknowledgment.

This work has received funding from the Horizon 2020 program of the European Union for research and innovation, under grant agreement no. 722176 (INDEED), EPSRC (grant nos. EP/P000916/1, EP/P000886/1, EP/P006973/1), and EPSRC National Epitaxy Facility.

References

[1] L. Balaghi, G. Bussone, R. Grifone, R. Hübner, J. Grenzer, M. Ghorbani-Asl, A. V. Krasheninnikov, H. Schneider, M. Helm, and E. Dimakis, Nat. Commun. **10**, 2793 (2019).

[2] M. Heurlin, D. Lindgren, K. Deppert, L. Samuelson, M. H. Magnusson, M. L. Ek, and R. Wallenberg, Nature **492**, 90 (2012)

[3] R. Yan, D. Gargas, and P. Yang, Nat. Photonics **3**, 569 (2009).

[4] T. Kuykendall, P. Ulrich, S. Aloni, and P. Yang, Nat. Mater. **6**, 951 (2007)

[5] Y. Liang, L. Zhai, X. Zhao, and D. Xu, J. Phys. Chem. B **109**, 7120 (2005).

[6] W. Guo, M. Zhang, A. Banerjee, and P. Bhattacharya, Nano Lett. **10**, 3356 (2010).

[7] C. P. T. Svensson, T. Mårtensson, J. Trägårdh, C. Larsson, M. Rask, D. Hessman, L. Samuelson, and J. Ohlsson, Nanotechnology **19**, 305201 (2008).

[8] S. W. Eaton, A. Fu, A. B. Wong, C. Z. Ning, and P. Yang, *Semiconductor Nanowire Lasers*, Nat. Rev. Mater. **1**, 16028 (2016)

[9] E. C. Garnett, M. L. Brongersma, Y. Cui, and M. D. McGehee, Annu. Rev. Mater. Res. **41**, 269 (2011).

[10] P. Krogstrup, H. I. Jørgensen, M. Heiss, O. Demichel, J. V. Holm, M. Aagesen, J. Nygard, and A. Fontcuberta I Morral, Nat. Photonics **7**, 306 (2013).

[11] A. I. Hochbaum and P. Yang, Chem. Rev. **110**, 527 (2010).

[12] D. Commandeur, G. Brown, E. Hills, J. Spencer, and Q. Chen, ACS Appl. Nano Mater. **2**, 1570 (2019).

[13] X. Long, F. Li, L. Gao, Y. Hu, H. Hu, J. Jin, and J. Ma, ChemSusChem **11**, 4094 (2018).

[14] D. Li, Y. Wu, P. Kim, L. Shi, P. Yang, and A. Majumdar, Appl. Phys. Lett. **83**, 2934 (2003).

[15] Y. S. Ju, Appl. Phys. Lett. **87**, 1 (2005).

[16] M. Y. Swinkels, M. R. Van Delft, D. S. Oliveira, A. Cavalli, I. Zardo, R. W. Van Der Heijden, and E. P. A. M. Bakkers, Nanotechnology **26**, (2015).

[17] J. M. Ziman, *Electrons and Phonons: The Theory of Transport Phenomena in Solids* (University Press, 1967).

[18] Y. He and G. Galli, Phys. Rev. Lett. **108**, 215901 (2012).

[19] P. Martin, Z. Aksamija, E. Pop, and U. Ravaioli, Phys. Rev. Lett. **102**, 125503 (2009).

[20] J. Shah, *Ultrafast Spectroscopy of Semiconductors and Semiconductor Nanostructures*, 2nd ed. (Springer-Verlag Berlin Heidelberg, 1999).

[21] L. Wittenbecher, E. Viñas Boström, J. Vogelsang, S. Lehman, K. A. Dick, C. Verdozzi, D. Zigmantas, and A. Mikkelsen, ACS Nano **15**, 1133 (2021).

[22] M. Willatzen, A. Uskov, J. Mork, H. Olesen, B. Tromborg, and A.-. Jauho, IEEE Photonics Technol. Lett. **3**, 606 (1991).

[23] M. Khodr, Res. J. Appl. Sci. Eng. Technol. **10**, 1384 (2015).

[24] J. Yan, M.-H. Kim, J. A. Elle, A. B. Sushkov, G. S. Jenkins, H. M. Milchberg, M. S. Fuhrer, and H. D. Drew, Nat. Nanotechnol. **7**, 472 (2012).

[25] D. K. Ferry, S. M. Goodnick, V. R. Whiteside, and I. R. Sellers, J. Appl. Phys. **128**, 220903 (2020).

[26] D. T. Nguyen, L. Lombez, F. Gibelli, S. Boyer-Richard, A. Le Corre, O. Durand, and J. F. Guillemoles, Nat. Energy **3**, 236 (2018).

[27] A. J. Nozik, Annu. Rev. Phys. Chem. **52**, 193 (2001).

[28] D. Tedeschi, M. De Luca, H. A. Fonseka, Q. Gao, F. Mura, H. H. Tan, S. Rubini, F. Martelli, C. Jagadish, M. Capizzi, and A. Polimeni, Nano Lett. **16**, 3085 (2016).

[29] C. Keong Yong, J. Wong-Leung, H. J. Joyce, J. Lloyd-Hughes, Q. Gao, H. Hoe Tan, C. Jagadish, M. B. Johnston, and L. M. Herz, Nano Lett. **13**, 4280 (2013).

[30] Y. Wang, H. E. Jackson, L. M. Smith, T. Burgess, S. Paiman, Q. Gao, H. H. Tan, and C. Jagadish, Nano Lett. **14**, 7153 (2014).

[31] Y. Zhang, A. M. Sanchez, Y. Sun, J. Wu, M. Aagesen, S. Huo, D. Kim, P. Jurczak, X. Xu, and H. Liu, Nano Lett. **16**, 1237 (2016).

[32] D. Catone, L. Di Mario, F. Martelli, P. O'Keeffe, A. Paladini, J. S. P. Cresi, A. K. Sivan, L. Tian, F. Toschi, and S. Turchini, Nanotechnology **32**, 025703 (2020).

[33] L. Tian, L. Di Mario, A. K. Sivan, D. Catone, P. O'Keeffe, A. Paladini, S. Turchini, and F. Martelli, Nanotechnology **30**, 214001 (2019).

[34] L. Tian, L. Di Mario, V. Zannier, D. Catone, S. Colonna, P. O'Keeffe, S. Turchini, N. Zema, S. Rubini, and F. Martelli, Phys. Rev. B **94**, 165442 (2016).

[35] H. J. Hovel, *Semiconductors and Semimetals. Volume 11. Solar Cells* (Academic Press, Inc.,New York, United States, 1975).

[36] D. E. Aspnes and A. A. Studna, Phys. Rev. B **27**, 985 (1983).

[37] N. Begum, M. Piccin, F. Jabeen, G. Bais, S. Rubini, F. Martelli, and A. S. Bhatti, J. Appl. Phys. **104**, 104311 (2008).

[38] A. G. Thompson, M. Cardona, K. L. Shaklee, and J. C. Woolley, Phys. Rev. **146**, 601 (1966).

[39] M. G. Craford, R. W. Shaw, A. H. Herzog, and W. O. Groves, J. Appl. Phys. **43**, 4075 (1972).

[40] M. Capizzi, S. Modesti, F. Martelli, and A. Frova, Solid State Commun. **39**, 333 (1981).

[41] J. Kübler, *O. Madelung (Ed.): Semiconductors (Group IV Elements and III—V*

*Compounds), Data in Science and Technology*, Springer-Verlag, Berlin 1991.

[42] D. M. Riffe, J. Opt. Soc. Am. B **19**, 1092 (2002).

[43] Y. Yang, D. P. Ostrowski, R. M. France, K. Zhu, J. Van De Lagemaat, J. M. Luther, and M. C. Beard, Nat. Photon. **10**, 53 (2016).

[44] D. Zanato, N. Balkan, B. K. Ridley, G. Hill, and W. J. Schaff, Semicond. Sci. Technol. **19**, 1024 (2004).

[45] O. Pagès, J. Souhabi, A. V. Postnikov, and A. Chafi, Phys. Rev. B **80**, 035204 (2009).

[46] M. E. Pistol and X. Liu, Phys. Rev. B **45**, 4312 (1992).

[47] F. Vallée, Phys. Rev. B **49**, 2460 (1994).

[48] H. M. Van Driel, Physical Review B **19**, 5928 (1979).

[49] W. Pötz and P. Kocevar, Phys. Rev. B **28**, 7040 (1983).

[50] J. Yang, X. Wen, H. Xia, R. Sheng, Q. Ma, J. Kim, P. Tapping, T. Harada, T. W. Kee, F. Huang, Y. B. Cheng, M. Green, A. Ho-Baillie, S. Huang, S. Shrestha, R. Patterson, and G. Conibeer, Nat. Commun. **8**, 14120 (2017).

[51] H. Shi, X. Zhang, X. Sun, and X. Zhang, Appl. Phys. Lett. **116**, 151902 (2020).

[52] Y. Rosenwaks, M. C. Hanna, D. H. Levi, D. M. Szmyd, R. K. Ahrenkiel, and A. J. Nozik, Phys. Rev. B **48**, 14675 (1993).

[53] V. B. Campos, S. Das Sarma, and M. A. Stroscio, Phys. Rev. B **46**, 3849 (1992).

[54] R. Cingolani, H. Lage, L. Tapfer, H. Kalt, D. Heitmann, and K. Ploog, Phys. Rev. Lett. **67**, 891 (1991).

[55] W. Pötz, Phys. Rev. B **36**, 5016 (1987).

[56] D. Catone, G. Ammirati, P. O'Keeffe, F. Martelli, L. Di Mario, S. Turchini, A. Paladini, F.

Toschi, A. Agresti, S. Pescetelli, and A. Di Carlo, Energies **14**, 708 (2021).